\documentclass[aps,pre,floatfix,superscriptaddress,twocolumn,twoside,a4paper,showpacs,showkeys,longbibliography,nofootinbib]{revtex4-2}

\newcommand\versionName{\texttt{JV: version of 2020.12.18---as accepted by RRR}}
\usepackage{graphicx}
\usepackage{amsmath,amssymb,amsfonts}
\usepackage{mathtools}
\usepackage{bm}

\usepackage{url}
\usepackage{hyperref}
\usepackage{color}

\usepackage{wasysym}

\newcommand{\ie}{{\textit{i.\,e.\@}}}
\newcommand{\eg}{{\textit{e.\,g.\@}}}
\newcommand{\cf}{{\textit{cf.\@}}}

\renewcommand{\vec}{\bm}

\newcommand{\rmd}{\ensuremath{\mathrm{d}}}

\usepackage{xcolor}

\newcommand{\zMSD}{\ensuremath{\eta}}
\newcommand{\moment}{\ensuremath{p}}

\newcommand{\zSM}{\alpha}
\newcommand{\zLW}{\beta_{\textrm{\tiny LW}}}
\newcommand{\zLLg}{\beta_{\textrm{\tiny LLg}}}

\newcommand{\zg}{\gamma}
\newcommand\Eq[1]{Eq.~(\ref{eq:#1})}
\newcommand\Eqs[1]{Eqs.~(\ref{eq:#1})}
\newcommand\EQ[1]{Equation~(\ref{eq:#1})}

\newcommand\Fig[1]{Fig.~\ref{fig:#1}}

\begin{document}
\preprint{\versionName}

\date{\today}

\title{Displacement Autocorrelation Functions for Strong Anomalous Diffusion:\\
  A Scaling Form, Universal Behavior, and Corrections to Scaling
}

\author{J\"urgen Vollmer}
\thanks{ORCID: \href{https://orcid.org/0000-0002-8135-1544}{0000-0002-8135-1544}}
\affiliation{\text{
    Institut f\"ur Theoretische Physik,
    Universit\"at Leipzig,
    Br\"uderstr.~16,
    D-04103 Leipzig, Germany}}
\affiliation{\text{
    Department of Mathematical Sciences, 
    Politecnico di Torino, 
    Corso Duca degli Abruzzi 24, 
    I-10129 Torino, Italy}}

\author{Lamberto Rondoni}
\thanks{ORCID:  \href{https://orcid.org/0000-0002-4223-6279}{0000-0002-4223-6279}}
\affiliation{\text{
    Department of Mathematical Sciences, 
    Politecnico di Torino, 
    Corso Duca degli Abruzzi 24, 
    I-10129 Torino, Italy}}
\affiliation{\text{
    INFN,
    Sezione di Torino,
    Via P. Giuria 1,
    10125 Torino, Italy}}
\affiliation{\text{
    Malaysia-Italy Centre of Excellence for Mathematical Sciences,
    Universiti Putra Malaysia,}
    \text{43400 Seri Kembangan,
    Selangor, Malaysia }}

\author{Muhammad Tayyab}
\affiliation{\text{
    Faculty of Engineering Sciences,
    Ghulam Ishaq Khan Institute of Engineering Sciences and Technology,}
    \text{Topi-2346, District Swabi, Khyber Pakhtunkhwa, Pakistan}}

\author{Claudio Giberti}
\thanks{ORCID: \href{https://orcid.org/0000-0002-5788-3903}{0000-0002-5788-3903}}
\affiliation{\text{
    Dipartimento di Scienze e Metodi dell'Ingegneria,
    Universit\`a di Modena e Reggio E., 
    Via Amendola 2, 
    I-42100 Reggio E., Italy}}

\author{Carlos Mej\'{i}a-Monasterio}
\thanks{ORCID: \href{https://orcid.org/0000-0002-6469-9020}{0000-0002-6469-9020}}
\affiliation{\text{
    Department of Mathematical Sciences, 
    Politecnico di Torino, 
    Corso Duca degli Abruzzi 24, 
    I-10129 Torino, Italy}}
\affiliation{\text{
    Laboratory of Physical Properties,
    Technical University of Madrid,
    Av. Complutense s/n,
    28040 Madrid, Spain}}

\begin{abstract}
  Strong anomalous diffusion is characterized by asymptotic power-law growth of the moments of displacement,
  with exponents that do not depend linearly on the order of the moment.
  The exponents concerning small-order moments are dominated by random motion,
  while higher-order exponents grow by faster trajectories, such as ballistic excursions or ``light fronts''.
  Often such a situation is characterized by two linear dependencies of the exponents on their order.
  Here, we introduce a simple exactly solvable model, the Fly-and-Die (FnD) model,
  that sheds light on this behavior and on the consequences of light fronts on displacement autocorrelation functions
  in transport processes.
  We present analytical expressions for the moments and derive a scaling form
  that expresses the long-time asymptotics of the autocorrelation function $\langle x(t_1)\,x(t_2)\rangle$ 
  in terms of the dimensionless time difference $(t_2-t_1)/t_1$. 
  The scaling form provides a faithful collapse of numerical data for vastly different systems. 
  This is demonstrated here for
  the Lorentz gas with infinite horizon,
  polygonal billiards with finite and infinite horizon, 
  the L\'evy-Lorentz gas,
  the Slicer Map,
  and L\'evy walks.
  Our analysis also captures the system-specific corrections to scaling.
\end{abstract}

\keywords{Anomalous transport, moments of displacement, displacement autocorrelation function}
\pacs{{05.60.-k, 66.30.je, 05.40.Fb, 05.45.-a}}

\maketitle
\markboth{\ \hfill J\"urgen Vollmer, Lamberto Rondoni, Muhammad Tayyab, Claudio Giberti, and Carlos Mej\'{i}a-Monasterio}%
{Displacement Autocorrelation Functions for Strong Anomalous Diffusion: A Scaling Form, \dots \hfill \ }

\section{Introduction}
\label{sec:intro}

A transport process  $x(t)$  is called anomalous 
if its mean-square displacement,
does not grow linearly in time~\cite{KRS08,HBA02,Zas02}.
Anomalous transport is found in a broad set of phenomena including
molecules moving in a living cell~\cite{Saxton},
dynamics on cell membranes~\cite{NHB07},
solid-state disordered systems~\cite{BG90},
telomeres inside the nucleus of mammalian cells~\cite{Bronstein2009},
soil transport~\cite{Rina},
heat transport in low-dimensional systems~\cite{LLP}.
Much work has been devoted to explore its microscopic origin~\cite{JBS03,KRS08,JBR08,Kla06,IRBP11,Metzler,Sok12,MJCB14,FSBZ15}.

The statistical properties of transport are given by the probability density of the displacement $P(\Delta x,t)$.
However, in most situations this density is unknown and transport is studied in terms of the asymptotic
behavior of its moments in time
\begin{equation} \label{eq:spectrum}
  \left\langle \bigl| \Delta x(t) \bigr|^{^\moment} \right\rangle
  \sim  t^{\zg(\moment)} \ ,
\end{equation}
where
$\langle \cdot \rangle$ denotes the average over an ensemble of trajectories,
$\moment \in \mathbb{R}$ is the order of the moment, and
the function~$\zg(\moment)$ is called the spectrum of the moments of the displacement.

The exponent $\zMSD=\zg(2)$ characterizes the mean-square displacement.
Transport is called
subdiffusive for $0 < \zMSD < 1$,
diffusive for $\zMSD = 1$, 
superdiffusive for $1 < \zMSD < 2$,
and ballistic when $\zMSD = 2$. 

Transport is termed scale-invariant
when the probability density of displacements $\Delta x$ admits a scaling
$P(\Delta x,t) = t^{-\nu}\mathcal{F}(\Delta x/t^\nu)$ with  $\nu$ constant,
\ie~when all moments of the displacement are characterized by a single scale $t^\nu$.
The spectrum of the moments of the displacement will then be a linear function of $\moment$:
$\zg(\moment)=\nu\, \moment$.

When the spectrum $\zg(\moment)$ is non-linear,
transport is called \emph{strong anomalous diffusion}~\cite{CMMGA99}.
This phenomenon has been observed in a variety of simple stochastic systems~\cite{ACMV00},
Hamiltonian systems with mixed phase space~\cite{CMMGA99},
polygonal billiard channels~\cite{SL06,JR06},
billiards with infinite horizon~\cite{AHO03,SS06,CESZ08},
one-dimensional maps~\cite{Pik91,SRGK15} and 
intermittent maps~\cite{CMMGA99,AC03},
running sand piles~\cite{CLNZ99},
stochastic models of inhomogeneous media~\cite{BCV10,BVLV14},
the diffusion in laser-cooled atoms~\cite{AKB17},
in experiments on the mobility of particles inside living cancer cells~\cite{GW10},
particles passively advected by dynamical membranes~\cite{SJ94}, and
in the bulk-mediated diffusion on lipid bilayers~\cite{KCNP16}.
A widely investigated case is the one in which 
different scalings of the bulk and the tail of the probability distribution
give rise to a piecewise linear form of the scaling exponent $\zg(\moment)$,
\begin{equation} \label{eq:SAD}
  \zg (\moment) =
  \left\{
    \begin{array}{lll}
      \nu \, \moment            & \quad\text{for \ } & \moment < \moment_c \, , \\
      \moment - (1-\nu) \, \moment_c  & \quad\text{for \ } & \moment \geq \moment_c \, .
    \end{array}
  \right .
\end{equation}

Strong anomalous diffusion is believed to be generic \cite{RDHB14,ZDK15,VBB18,LastBurioniArXiv}
for dynamics with fat-tailed waiting-time distributions.
In recent years its dynamical basis has been analyzed through generalizations of the central limit theorem
and non-normalizable densities \cite{RDHB14,RDHB14b,FSBZ15,LastBurioniArXiv}.
At the same time, numerous studies have addressed the relation between the properties of deterministic dynamics and transport.
Roughly speaking, chaos is commonly associated with fast decay of correlations and normal diffusion,
while non-chaotic dynamics often leads to anomalous transport and slow decay of correlations \cite{SRGK15,Zas02,Kla06,KRS08,JR06}.
Stochastic processes, on the other hand, are often considered to resemble chaotic dynamics,
but they can give rise to normal as well as anomalous diffusion,
depending on their correlations decay rate~\cite{DC00,CFVN02}.
Thus, numerous questions remain open \cite{Zas02,Kla06,DKU03,LWWZ05,Sok12,BF99,BFK00,BCV10}.
In particular, the asymptotic behavior of correlation functions is not understood in general,
although it is relevant \eg~to distinguish transport processes
that arise from different microscopic mechanisms,
but have the same moments \cite{Sok12,BF07,BS07,Zab08,ZDK15}.

In~\cite{SRGK15} a deterministic map named Slicer Map (SM) was introduced to shed light on these issues.
It was shown that for an appropriate matching of parameters its moments of the displacement, of order $2$ and higher,
scale in time like those of the L\'evy-Lorentz gas (LLg)~\cite{BCV10},
a transport model~\cite{SRGK15,GRTV17} where strong anomalous diffusion emerges
as a consequence of scattering in an environment with quenched disorder.
In \cite{GRTV17} it was found that the matching of exponents also
entails the equality of the large-time scaling of the displacement autocorrelation function,
\begin{equation}\label{eq:phiDef}
  \phi(t_1, t_2) = \bigl\langle \Delta x(t_1) \: \Delta x(t_2) \bigr\rangle \, ,
\end{equation}
making two entirely different systems hardly distinguishable on the level of the statistics of displacements.

Here, we introduce the Fly-and-Die (FnD) model
as a minimal exactly solvable deterministic model where strong anomalous diffusion emerges due to a light front,
\ie~ballistic trajectories that did not undergo transitions up to any finite time $t$. 
Unlike the SM the FnD is continuous in time and space.
Otherwise, it closely mimics the asymptotic transport properties of the SM (and hence of the LLg).
We present analytic expressions for its spectrum of the moments $\zg(\moment)$
and its displacement autocorrelation function $\phi(t, t+h)$.
Our main result is a scaling form expressing the autocorrelation function
as function of the reduced time $h/t$, 
\begin{align}
  \label{eq:scaling-form}
  \frac{  \phi(t, t+h) }{  \phi(t, t) } - 1 = C\!\left( \frac{h}{t} \right) \, .
\end{align}
This result is interesting for
two reasons:
i.~There are only very few analytic results about position-position autocorrelation function for strong anomalous diffusion.
To our knowledge they have only been calculated for L\'evy walks~\cite{FB13a,FB13b}
and the slicer map~\cite{GRTV17}, 
and those studies did not introduce the scaling form,  \Eq{scaling-form}.
ii.~We propose that \Eq{scaling-form} is universally valid.
It applies for every system showing strong anomalous diffusion where the mean-square displacement is governed by the light front.
This claim is underpinned in the second part of the paper by showing
that the correspondence between the transport properties of the FnD, the SM, and the LLg extends to a much broader class of systems. 
To this end we compare the analytical dependencies of the FnD with the scaling function for L\'evy walks
that is derived here based on analytical results of Ref.~\cite{FB13b},
and with numerical results for systems where analytical treatments are beyond reach.
For all dynamics the scaling form, \Eq{scaling-form}, provides a very good data collapse.
However, careful inspection reveals
that there are small systematic corrections to scaling
with system-specific features,
or related to non-asymptotic effects.

This paper is organized as follows:
In Section~\ref{sec:FnD}, we introduce the FnD dynamics,
and obtain analytical expressions for the spectrum of moments, 
and the autocorrelation function.
In Section~\ref{sec:moments}, we revisit the
Slicer Map (Sec.~\ref{sec:SM}),
L\'evy Walks (Sec.~\ref{sec:LW}),
the L\'evy Lorentz gas (Sec.~\ref{sec:LLg}),
the Lorentz gas (Sec.~\ref{sec:LG}), and 
a family of polygonal billiard channels (Sec.~\ref{sec:PBC}),
comparing their spectrum of moments, $\zg(\moment)$, with that of the FnD dynamics.
The scope of this discussion is strictly confined to material required to address the correlation function in a self-contained manner.
We provide links to the literature for relevant further work.
In Section~\ref{sec:correlations}, the comparison concerns the scaling of the autocorrelation function $\phi(t_1,t_2)$,
which is not treated elsewhere.
Section~\ref{sec:discussion} highlights predictions of the FnD model that apply universally for all models
from the perspective of scaling theory and the Buckingham-Pi theorem,
and it addresses model-specific features.
Section~\ref{sec:conclusion} summarizes our findings.

\section{The Fly-and-Die Model}
\label{sec:FnD}

In this section we introduce the Fly-and-Die (FnD) model.

\subsection{Dynamics}
\label{sec:FnD-dynamics}

The FnD dynamics addresses the motion of a particle on the semi-infinite line $[0,\infty)$.
Different trajectories are labeled by initial conditions $x_0 \in [0,1]$.
Starting at $x_0$ the particle initially moves along the positive $x$ axis with unit velocity (it \emph{flies}). 
At a time $t_c(x_0)$ it stops and never resumes motion (it \emph{dies}). 
Hence, at time $t$ a particle is found in position
\begin{subequations}\label{eq:FnD-EOM}
\begin{equation}
  x( x_0, t ) = \left\{
  \begin{array}{l@{\quad\text{for \ }}l}
    x_0 + t          & t \leq t_c(x_0) \,, \\
    x_0 + t_c( x_0 ) & t   >  t_c(x_0) \,.
  \end{array}
  \right .
\end{equation}
Anomalous transport emerges
when the distribution of the flight times $t_c(x_0)$ has a power-law tail.
Here, we consider the case 
\begin{equation}\label{eq:tcx0}
  t_c(x_0) = \left( \frac{b}{ x_0 } \right)^{1/\xi} \, ,
\end{equation}
\end{subequations}
with $\xi$ and $b$ positive constants, and initial conditions~$x_0$
uniformly distributed in the interval $[0, 1]$.
The minimum flight time is $t_M = b^{1/\xi}$.

In simple words, the FnD dynamics evolves an ensemble of initial conditions in the unit interval.
Each point is associated to a trajectory that flies independently and without turning back for a time
that is determined by the initial condition $x_0$ via the function $t_c(x_0)$.
The distribution of the flight times $t_c$ determines the transport properties of the FnD dynamics.

\subsection{Moments of displacement}

For the FnD dynamics the spectrum of the moments of the displacement is obtained straightforwardly:
Initial conditions, $x_0$, that perform a flight longer than $t$ lie in the interval $[0, (t_M/t)^\xi)$.
Therefore, adopting the uniform distribution of initial conditions in the unit interval,
the probability to fly for a time larger than $t$ is given by
\begin{equation}
  P(>t) = \frac{ b }{ t^{\xi}  } \, ,
  \qquad \text{for \ }
  t \geq t_M 
\end{equation}
In the following we only consider times $t \geq t_M$.

For $\moment \neq \xi$ the $\moment^{\text{th}}$ moment
of the displacement $\Delta x(t)$ takes the form
\begin{subequations}\label{eq:FnD-moments}
\begin{align}
  \langle |\Delta x(t)|^\moment \rangle 
  &=
  \langle | x( x_0, t ) - x_0 |^\moment \rangle 
\nonumber\\[2mm]
  &=
  \int_0^1 \rmd x_0 \: | x( x_0, t ) - x_0 |^\moment
\nonumber\\[2mm]
  &=
    \int_0^{P(>t)} \rmd x_0 \, t^\moment
  + \int^1_{P(>t)} \rmd x_0 \, \left( t_c(x_0) \right)^\moment
\nonumber\\[2mm]
  &=
  t_M^\xi \, t^{\moment - \xi} + \frac{ \xi \, t_M^{\moment}}{\xi - \moment} \: \left[ 1 - \left( \frac{t}{t_M} \right)^{\moment - \xi} \right]
\nonumber\\[2mm]
  &=
    \frac{ t_M^\xi \: t^{\moment - \xi}}{\moment - \xi} \; \left[ \moment - \xi \: \left( \frac{t_M}{t} \right)^{\moment - \xi} \right]
    \, ,\text{ \ for }
    \moment \neq \xi
    \, .
\label{eq:FnD-moments-p-neq-xi}
\end{align}
For $\moment = \xi$ an analogous calculation yields
\begin{align}
  \langle |\Delta x(t)|^\moment \rangle 
  &=
    \int_0^{P(>t)} \rmd x_0 \, t^\moment
  + \int^1_{P(>t)} \rmd x_0 \, \frac{t_M^\xi}{x_0}
  \nonumber\\[2mm]
  &=
    t_M^\xi \, t^{\moment - \xi} + t_M^\xi \, \ln \left( \frac{t}{t_M} \right)^\xi
    \nonumber\\[2mm]
  &=
    t_M^\xi \, \left( 1 + \xi \: \ln \frac{t}{t_M} \right)
    \, , \qquad \text{ \ for }
    \moment = \xi
    \, .
    \label{eq:FnD-moments-p-eq-xi}
\end{align}
This expression can also be found as the $\moment \to \xi$ limit of \Eq{FnD-moments-p-neq-xi},
by observing that
$(t_M/t)^{\moment-\xi} = \exp(- (\moment-\xi) \: \ln(t/t_M))
                     = 1 - (\moment-\xi) \: \ln(t/t_M)$
up to corrections of order $(\moment-\xi)^2$.

At large times, $t \gg b^{1/\xi}$, the spectrum of the moments takes the form:
\begin{equation}\label{eq:FnD-moment-scaling}
  \langle | \Delta x(t) |^\moment \rangle
  \simeq
  \left\{
  \begin{array}{l@{\quad\text{for \ }}l}
    \frac{\xi}{\xi - \moment} \: t_M^{\moment}
    & \moment < \xi \, ,
    \\[2mm]
    \xi \, t_M^\xi \: \ln (t / t_M )
    & \moment = \xi \, ,
    \\[2mm]
    \frac{\moment}{\moment - \xi} \: t_M^\xi \: t^{\moment - \xi}
    & \moment > \xi \, .
  \end{array}
  \right .
\end{equation}
\end{subequations}
The FnD exhibits piecewise-linear scaling in the statistics of the displacement
with an exponent $\zg(\moment)$
that is zero for $\moment <\xi$
and linear, $\zg(\moment) = \moment-\xi$, for $\moment > \xi$.
The mean-square displacement, $\langle | \Delta x |^2 \rangle$,
approaches a constant value (localization) for $\xi  > 2$,
it grows logarithmically for $\xi = 2$, and
it grows as a power law with a mean-square displacement scaling exponent $\zMSD  = 2-\xi$ for $0< \xi < 2$.
Then,
$\xi \in (0  , 1)$ corresponds to  superdiffusive behavior,
$\xi  =   1$       to normal diffusion, and
$\xi \in (1 , 2)$  to subdiffusion:
the FnD gives rise to all anomalous transport regimes.

\subsection{Displacement autocorrelation function}

We now turn our attention to the autocorrelation function 
\begin{align}\nonumber
  \phi(t_1, t_2) 
  &:=
  \langle \left( x( x_0, t_1 ) - x_0 \right) \: \left( x( x_0, t_2 ) - x_0 \right) \rangle 
\\[2mm]
  &=
  \int_0^1 \rmd x_0 \: \left( x( x_0, t_1 ) - x_0 \right) \: \left( x( x_0, t_2 ) - x_0 \right) \, .
\nonumber
\end{align}
Without loss of generality, we set $t_2 > t_1$.  Accordingly, we split
the integration range in three intervals
\begin{description}
\item[$0 < x_0 < P(>t_2)$ ]
  The trajectory is still flying at time~$t_2$,
  hence $\Delta x(t_1) = t_1$ and $\Delta x(t_2) = t_2$. 
\item[$P(>t_2) < x_0 < P(>t_1)$ ] 
  The trajectory was still flying at time $t_1$ but it died by the time $t_2$.
  Consequently, $\Delta x(t_1) = t_1$ and $\Delta x(t_2) = t_c(x_0)$.
\item[$P(>t_1) < x_0 < 1$ ] 
  The trajectory died before $t_1$.
  Consequently, $\Delta x(t_1) = \Delta x(t_2) = t_c(x_0)$.
\end{description}
Performing calculations analogous to the derivation of \Eq{FnD-moments-p-neq-xi},
one finds for $\eta > 0$ that
\begin{equation}\label{eq:FnD-v-correlation}
  \phi(t_1, t_2) 
  = 
  \frac{t_M^{2-\zMSD} \: t_1 \: t_2^{\zMSD - 1} }{\zMSD-1} 
  - \frac{(2-\zMSD) \: t_M^{2-\zMSD} \, t_1^\zMSD }{\zMSD \, (\zMSD-1)} 
  - \frac{(2-\zMSD) \: t_M^2}{\zMSD} \, .
\end{equation}
It is convenient to normalize this expression by
\begin{equation}\label{eq:FnD-mean-square}
  \phi(t_1, t_1)
  = \langle | \Delta x(t_1) |^2 \rangle
  =
  \frac{2 t_M^{2-\zMSD} \: t_1^{\zMSD}}{\zMSD} \; \left[ 1 - \frac{2 - \zMSD}{2} \: \left(\frac{t_M}{t_1} \right)^{\zMSD} \right] \, .
\end{equation}
Subtracting one and denoting the time difference by $h = t_2 - t_1$,
one obtains
\begin{align}\label{eq:FnD-correlation-ratio}
  C\!\left( \frac{h}{t_1} \right)
  &=
  \frac{ \phi(t_1, t_2) }{ \phi(t_1, t_1) } - 1
  = \frac{1}{2} \: \frac{\eta}{\eta-1} \;
    \frac{  \left( 1 + \frac{ h }{ t_1 } \right)^{\eta-1} - 1 }
    { 1 - \frac{2 - \zMSD}{2} \: \left(\frac{t_M}{t_1} \right)^{\zMSD} }
\end{align}
In the large-time limit $t_M/t_1 \to 0$ for a fixed value of $h/t_1$,
\ie~for a finite ratio $t_2/t_1$,
this entails a scaling form
\begin{subequations}\label{eq:FnD-correlation-scalingForm}
  \begin{align}\label{eq:FnD-correlation-scaling}
    C\!\left( \frac{h}{t_1} \right) 
    &= \left\{
      \begin{array}{lll}
        \frac{1}{2} \: \frac{\eta}{\eta-1} \;
          \left[ \left( 1 + \frac{ h }{ t_1 } \right)^{\eta-1} - 1 \right]
        & \text{ for } & \zMSD \ne 1 \, ,
        \\[3mm]
        \frac{1}{2} \: \ln\left( 1 + \frac{h}{t_1} \right)
        & \text{ for } & \zMSD = 1 \, .
      \end{array} \right.
  \end{align}
  Here, again the $\zMSD = 1$ case is found as the $\zMSD \to 1$ limit of the $\zMSD\ne 1$ case.
  \Eq{FnD-correlation-scaling} connects the large-time behavior of the correlation function to the dependence of the mean-square displacement.
  It has the following asymptotic scaling for small and large values of $h/t_1$,
  \begin{equation}\label{eq:FnD-correlation-asymptotics}
    C\!\left( \frac{h}{t_1} \right) 
    \simeq \left\{
      \begin{array}{l@{\;\;\text{for \ }}l}
        \frac{\eta}{2} \: \frac{h}{t_1}  
        & \frac{h}{t_1} \ll 1 \, ,
        \\[2mm]
        \frac{1}{2} \: \frac{\zMSD}{\zMSD-1} \: \left( \frac{h}{t_1} \right)^{\zMSD-1}
        & \frac{h}{t_1} \gg 1 \, ,  \; \zMSD > 1 \, ,
        \\[2mm]
        \frac{1}{2} \; \ln\!\left( \frac{h}{t_1} \right)
        & \frac{h}{t_1} \gg 1 \, ,  \; \zMSD = 1  \, ,
        \\[2mm]
        \text{const}
        & \frac{h}{t_1} \gg 1 \, , \; \zMSD < 1 \, .
      \end{array} 
    \right .
  \end{equation}
\end{subequations}
As anticipated in \Eq{scaling-form}
the scaling form, \Eq{FnD-correlation-scalingForm},
represents the time dependence of the displacement autocorrelation function $\phi(t_1, t_2)$
in terms of a single dimensionless time ratio $h/t_1 = (t_2-t_1)/t_1$.

\subsection{A remark on the equivalence of statistics}
\label{sec:FnD-equivalence}

Equation~\eqref{eq:FnD-moment-scaling} entails
that the moments of the FnD dynamics show a piecewise-linear behavior,
as in the case of strong anomalous diffusion.

For  $\moment > \xi$  the moments of the FnD dynamics grow in time with an exponent $\zg(\moment) = \moment-\xi$.
A linear behavior with slope one for $\zg(\moment)$ is typical
when the contributions to the higher cumulants are asymptotically dominated by ballistic flights
or light fronts~\cite{CMMGA99,VBB18,LastBurioniArXiv}.
For L\'evy processes this was explicitly discussed in \cite{ZDK15}.
Recently, it has been suggested that the scaling holds in general due to a
``big jump principle''~\cite{VBB18,LastBurioniArXiv}.
The mean-square displacement of the FnD dynamics follows this behavior when $\xi < 2$.
In the next section we will explicitly verify
that the matching of the value of the mean-square displacement will then also fix all other moments
in the large-$\moment$ branch of the piecewise-linear behavior of~$\zg(\moment)$. 

For the FnD dynamics, the moments $\moment < \xi$ are constant in time.
This is a peculiarity of the dynamics that is non-generic:
the FnD dynamics does not mimic the lower moments of displacement for other dynamics.
This is a strong point of the FnD model.
It allows us to clearly pin down features of dynamics
that derive from rare ballistic flights.
The FnD prediction, \Eq{FnD-correlation-scalingForm},
for the displacement correlation function $\phi(t_1,t_2)$
is of particular interest.
It suggests a scaling form, \Eq{scaling-form},
that turns out to be astonishingly robust in vastly different settings.

\section{Models and Moments}
\label{sec:moments}

In this section, we investigate
how the spectrum of the moments of the displacement in dynamics with light fronts
can be mapped to the spectrum of the FnD.
First, we address the relation of the FnD to the Slicer Map (SM).
Then, we consider a L\'evy-walk model,
that is the most widely studied model featuring strong anomalous diffusion. 
Subsequently, we discuss 
a continuous-time stochastic process, the L\'evy-Lorentz gas (LLg), 
a chaotic billiard with infinite horizon, the periodic Lorentz Gas (LG), and
some Polygonal Billiard Channels (PBC).
For each system we also briefly revisit key features of its dynamics.

\subsection{The Slicer Map (SM)}
\label{sec:SM}

The SM is a one-parameter deterministic exactly solvable model
$S_\zSM : \left[  0 , 1 \right]  \times \mathbb{Z}
        \to \left[  0 , 1 \right] \times \mathbb{Z}$
defined by~\cite{SRGK15,GRTV17}:
\begin{align}\label{eq:SM-EOM}
  S_\zSM(x,m) := \nonumber \\
  &  \hskip -40pt  
    \left\{
    \begin{array}{rl}
      (x,m-1) & \mbox{ if } 0\leq x \le  \ell_{m} \mbox{ or } \frac{1}{2} < x \leq 1-\ell_{m},\\[2mm]
      (x,m+1) & \mbox{ if } \ell_{m} < x \leq \frac{1}{2} \mbox{ or } 1-\ell_{m} < x \leq 1 \, .
    \end{array}
                \right.
\end{align}
For all integers $m$ the  so-called ``\emph{slicer}'' $\ell_m$ 
is defined by
\begin{equation}
  \ell_m := \frac{1}{(|m|+2^{1/\zSM})^\zSM} \quad\text{ with } \zSM \in \mathbb{R}^+ \, .
\label{eq:SM-slicer}
\end{equation}
For $1/2 <  x < 1$ each iteration of the map increases the values of $m$ by one,
until $x > \ell_m$.
Subsequently, the trajectory enters a stable period-two cycle,
oscillating back and forth between the two neighboring sites $m$ and $m-1$.
Similarly, for $0 < x  < 1/2$ each iteration of the map decreases the values of  $m$  by one,
until $x  < -\ell_m$,  and
then the trajectory enters a stable period-two cycle.

The SM was inspired by the dynamics of polygonal billiards,
which have vanishing Lyapunov exponents,
because their trajectories separate substantially only at a countable set of points~\cite{JR06,SL06,JBR08}.
Analogously, the distance between two points $x_1$ and $x_2$ of the SM jumps  discontinuously
when they reach a cell $m$ where $\ell_m \in [x_1, x_2]$.
The corners of polygons act as slicers of the bundle of initial conditions
(\cf~the discussion of polygonal billiards in Section~\ref{sec:PBC}).

The moments of the displacement of an ensemble
where particles are initially distributed uniformly in the interval $[0,1]$
were obtained in~\cite{SRGK15},
\begin{subequations}
\begin{align} 
  \bigl\langle |x - x_0|^{^\moment} \bigr\rangle
  &= 2 \; \sum_{k=1}^{n-1} k^{^\moment} \: \Delta_k(\zSM) \: \bigl( 1 + \mathcal{O}( k^{-1} ) \bigr)
  \\[1mm]
  & + 2 \, n^{^\moment} \; \sum_{k=n}^\infty \Delta_k(\zSM)
\end{align}
where $\Delta_k(\zSM) = \ell_{k-1}(\zSM) - \ell_k(\zSM)$.
In the large-time limit, $n \gg 2^{1/\zSM}$, 
this expression has the asymptotic behavior
\begin{equation} \label{eq:SM-moments}
  \left\langle (x_n - x_0)^{^\moment} \right\rangle \simeq \left\{
    \begin{array}{l@{\quad\text{for \ }}l}
      \text{const}                                        & \moment < \zSM \, , \\[1mm]
      2 \; \ln\frac{n^\zSM}{2}                           & \moment = \zSM \, ,\\[1mm]
      \displaystyle \frac{2\, \moment}{\moment-\zSM} \, n^{p-\zSM}  & \moment > \zSM \, .
    \end{array}\right. 
\end{equation}
\end{subequations}
The SM exhibits strong anomalous  diffusion with a piecewise-linear spectrum $\zg(\moment)$ and threshold order $\moment_c=\zSM$.
The  SM exhibits all transport  regimes
upon varying the parameter  $\zSM$.
It features subdiffusion for $2  \ge  \zSM >  1$,
diffusion for $\zSM = 1$, and
superdiffusion for $1  >  \zSM >  0$.

Comparing \Eq{SM-moments} with \Eq{FnD-moment-scaling},
we see that the spectrum of the moments of the displacement of the FnD dynamics and the SM coincide
when taking $\xi=\zSM$ and $b=t_M^\xi=2$.
These expressions for $\xi$ and $b$ can also be found by matching the mean-square displacement, \ie~the expressions for $\moment = 2$,
and observing that the two branches of $\zg(\moment)$ have slope zero and one, respectively.

\subsection{L\'evy walks (LW)} 
\label{sec:LW}

L\'evy walks refer to dynamics
where an agent walks
with constant speed $v$ 
along straight line segments of length $r$
that are sampled from a probability distribution with a power-law tail
\cite{ACMV00,MJCB14,ZDK15}.
In the simplest case the dynamics is one-dimensional.
At the end of each segment there is a random decision taken to go left or right,
and the length $r$ is sampled from the probability distribution
\begin{equation}\label{eq:lambdaLW}
  \lambda(r) \equiv \frac{\zLW}{r_0}  \; \left( \frac{r}{r_0} \right)^{-\zLW-1}, \quad r\in[r_0,\infty),
\end{equation}
where the exponent $\zLW > 0$ characterizes the power-law decay of the distribution,
and $r_0>0$ is the minimum distance between scatterers.

The moments of displacement of an ensemble of particles
that start a LW at $x=0$ were obtained in \cite{ACMV00,RDHB14b},
\begin{align}\label{eq:LW-moments}
  \zg (\moment) =
  \left\{
    \begin{array}{lll}
      \moment             & \quad\text{for \ } & \zLW < 1 \, , \\[2mm]
      \moment / \zLW      & \quad\text{for \ } & \zLW > 1 \: \land \: 0 < \moment < \zLW  \, , \\[2mm]
      \moment + 1 - \zLW  & \quad\text{for \ } & \zLW > 1 \: \land \: \zLW < \moment \ \, .
    \end{array}
  \right .
\end{align}
For $\zLW < 1$ the LW exhibits anomalous diffusion with ballistic scaling, $\zMSD = 2$. 
For $\zLW > 1$ it features strong anomalous diffusion with a piecewise-linear spectrum $\zg(\moment)$
and threshold order $\moment_c=\zLW$.
For $\zLW > 2$ there is subdiffusion with exponent $\zMSD = 2/\zLW$,
and
for $1 < \zLW < 2$ superdiffusion with exponent $3-\zLW$. 

Comparing \Eq{LW-moments} with \Eq{FnD-moment-scaling},
we see
that the spectrum of the moments of the displacement of the FnD dynamics
and the LW coincide for large moments~$\moment$,
when one matches parameters as $\xi=-1+\zLW$.
In the superdiffusive regime this follows from matching the mean-square displacement,
and acknowledging that superdiffusion emerges in the models due to rare, very long ballistic trajectories,
\ie~that $\zg(\moment)$ has slope one.

\subsection{The L\'evy-Lorentz Gas (LLg)}
\label{sec:LLg}

The L\'evy-Lorentz gas (LLg) was introduced in~\cite{BF99} (and studied further in \eg~\cite{BCV10,BFK00})
as a one-dimensional model of anomalous transport in semiconductor devices.
In the LLg model a particle is randomly scattered with probability $1/2$ at randomly fixed positions on the line.
Between two consecutive collisions the particle moves at constant velocity $\pm v$.
The distances  $r$ between neighboring scatterers are sampled from a L\'evy distribution with probability density
\begin{equation}
  \lambda(r) \equiv \frac{\zLLg}{r_0}  \; \left( \frac{r}{r_0} \right)^{-\zLLg-1}, \quad r\in[r_0,\infty),
\end{equation}
\ie~the same probability distribution also adopted for the LW, \cf~\Eq{lambdaLW}.

The spectrum of the moments of the displacement was derived under the assumption
that higher moments are dominated by the light front,
an assumption also denoted as single long jump principle.
For non-equilibrium initial conditions,
\ie~an ensemble of trajectories starting all at the same scatterer position,
it entails~\cite{BCV10}
\begin{equation}\label{eq:LLg-moments}
  \zg(\moment) 
  =
  \left\{
    \begin{array}{l@{\quad\text{for \ }}l}
      {\frac{\moment}{1+\zLLg}}              & \zLLg  <   1,\ \moment <\zLLg    \, ,\\[1mm]
      {p-\frac{\zLLg^{2}}{1+\zLLg}}    & \zLLg  <   1,\ \moment >\zLLg    \, ,\\[1mm]
      {\frac{\moment}{2}}                    & \zLLg \geq 1,\ \moment <2\zLLg-1 \, ,\\[1mm]
      {\frac{1}{2}+p-\zLLg}            & \zLLg \geq 1,\ \moment >2\zLLg-1 \, . 
    \end{array}
  \right. 
\end{equation}
The LLg shows strong anomalous diffusion with mean-square displacement
$\langle r^{2}(t)  \rangle \sim  t^\zMSD$
growing as a power law in time with exponent
\begin{equation}\label{eq:LLg-MeanSquare}
  \zMSD 
  =
  \left\{
    \begin{array}{l@{\quad\text{for \ }}r@{\;\zLLg\;}l}
      2 - \frac{\zLLg^{2}}{1+\zLLg}   &               &  \le       1      \, ,\\[1mm]
      \frac{5}{2}-\zLLg               &       1      < & \le \frac{3}{2}  \, ,\\[1mm]
      1                               & \frac{3}{2}  < &                 \, .
    \end{array}
  \right. 
\end{equation}
It exhibits superdiffusive transport for  $0<\zLLg< 3/2$,
and diffusive transport for $\zLLg \ge 3/2$.
Unlike the FnD dynamics and the SM, it has no subdiffusive transport regime.
On the other hand, the LLg dynamics is much more complex as compared to the former models.
Only its moments of displacement, \Eq{LLg-moments}, have been analytically expressed. 

In~\cite{SRGK15} it was established
that the exponent $\zMSD$ characterizing the mean-square displacement of the LLg, \Eq{LLg-MeanSquare},
agrees with the relation $\zMSD = 2-\alpha$ for the SM dynamics,
if
\begin{equation} \label{eq:match}
  \alpha = \left\{
    \begin{array}{l@{\quad\text{for \ }}r@{\;\zLLg\;}l}
      {\frac{\zLLg^2}{1+\zLLg}}    &               & \le       1     \,, \\[1mm]
      \zLLg - \frac{1}{2}          & 1           < & \le  \frac{3}{2} \,, \\[1mm]
      1                            & \frac{3}{2} < &                 \,.
    \end{array}
  \right. 
\end{equation}
For $\zLLg < 3/2$ the mean-square displacement, $\moment = 2$, lies in one of the large-$\moment$ branches of $\zg(\moment)$,
such that the matching of the mean-square displacement entails an agreement of the spectrum of the moments of the displacement
for $\moment >\zLLg$ when $0<\zLLg<1$, 
and for $\moment >2\zLLg-1$ when $1<\zLLg<3/2$. 
On the other hand, for $\zLLg>3/2$ the matching only works for the mean-square displacement
because \Eq{FnD-moment-scaling} requires $\zg(\moment) = \moment-\xi = \moment-1$,
while \Eq{LLg-moments} stipulates $\zg(\moment) = \moment/2$.

We conclude
that the matching of the mean-square displacement 
extends to the large-$\moment$ branch of the spectrum $\zg(\moment)$ of the LLg
by taking  $\xi = \alpha$.  
The matching of the parameter $\alpha$ of the SM with $\zLLg$ of the LLg involves three branches, \Eq{match}, 
while there is a single linear relation required for the matching of the SM and the FnD dynamics.
The authors of \cite{SRGK15} took this non-trivial matching
as evidence that the equivalence is not complete.

\subsection{The Lorentz gas (LG)}
\label{sec:LG}

Billiards with infinite horizon are a class of deterministic dynamics
where the length of free flights is not bounded.
The resulting light fronts induce rare ballistic excursions of the displacement,
and hence strong anomalous diffusion~\cite{AHO03}.
A distinguished example is the Lorentz gas (LG),
which is a paradigmatic model of deterministic diffusion~\cite{Sinai80,D14}.

\begin{figure}[t]
  \centering
  \includegraphics[width=0.47\textwidth]{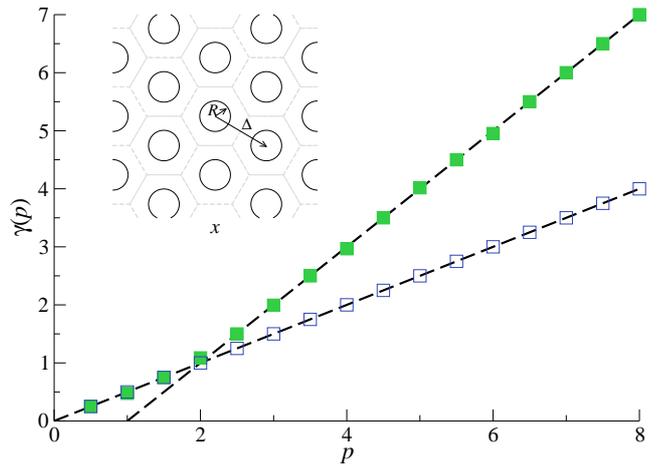}
  \caption{(Color online).
    Spectrum of the moments of displacement $\zg(\moment)$ for a LG billiard with
    infinite horizon, $R=0.1$ (green solid squares), and
    finite horizon, $R=1.1$ (blue open squares).
    The dashed curves correspond to the scaling given by the two linear functions in
    \Eq{spec-LG}.
    Inset: Schematic geometry of the LG. 
    The gray hexagonal lines are to guide the eye.
    \label{fig:LG-moments}}
\end{figure}

The periodic LG consists of an array of circular scatterers of radius $R$
periodically placed on a triangular lattice of the plane.
The separation between nearest-neighbor scatterers is set to $\Delta = \frac{8}{3}\cos\frac{\pi}{6}$
so that the horizon is infinite for $R<1$
(see inset of \Fig{LG-moments}).
The dynamics consists of free flights of point particles of unit velocity
that are started in a region close to the origin 
and undergo specular collisions with the scatterers.

With infinite horizon the variance of trajectory lengths diverges
because the probability of a trajectory segment with no collision
and length between $l$ and $l+dl$ decays as $l^{-3}$.
In a seminal work Bleher showed that for every periodic configuration of scatterers with infinite horizon
the displacement scaled by $\sqrt{t\ln(t)}$ converges in distribution to Gaussian statistics \cite{B92}.
This weak superdiffusion is hard to be observed numerically
as the time scales at which it sets in are physically unobservable.
Since Bleher's result the existence of this weak superdiffusion in infinite horizon billiard tables has generated a great deal of research \cite{C94,AHO03,D12,CGLS14b,MT16}.
Recently in Ref.~\cite{ZPFDB18} a refined central limit theorem
in which the $\sqrt{t\ln(t)}$ scaling to the displacement is substituted by a rescaled Lambert function has shown to appropriately describe the mean-square displacement,
at least for some geometries of the infinite horizon periodic LG (see also \cite{FouxonDitlevsen2019}).

At infinite horizon ($R<1$), it was observed that the LG exhibits strong anomalous diffusion~\cite{AHO03},
with spectrum
\begin{equation} \label{eq:spec-LG}
  \zg(\moment) = \left\{
    \begin{array}{l@{\quad\text{for \ }}l}
      \frac{\moment}{2}  &   \moment < 2 \,,\\
      \moment - 1        &   \moment > 2 \,.
    \end{array} \right.
\end{equation}
When the horizon is finite ($R\ge 1$), transport is diffusive,
and the moments of displacement scale in time with exponent $\zg(\moment) = \moment/2$ for any order $\moment$.
For $R<1$, the mean-square displacement is at the cross-over of the two branches of the spectrum of the moments of the displacement:
For moments $\moment < 2 $ the spectrum shows normal diffusive transport,
while it is dominated by the ballistic excursion for $\moment > 2$.
The latter branch agrees with the FnD dynamics with $\xi=1$. 

This is shown in \Fig{LG-moments},
where results are reported for numerical simulations of $10^6$ trajectories lasting a total time of~$10^6$,
with randomly chosen initial positions $\vec{r}(0)$ and velocities $\vec{v}(0)$.
Solid squares show the spectrum  $\zg(\moment)$ for the infinite-horizon case with $R=0.1$,
and open squares for the finite-horizon case with $R=1.1$.

\subsection{Polygonal billiard channels (PBC)} 
\label{sec:PBC}

\begin{figure}[t]
  \centering
  \includegraphics[width=0.47\textwidth]{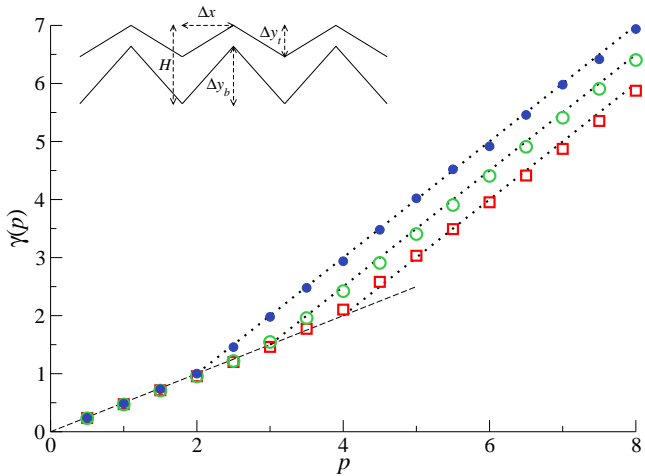}
  \caption{(Color online).
    Spectrum of  the moments of displacement $\zg(\moment)$  for
    the polygonal  channel  with
    $\Delta x = 1$,
    $\Delta y_t = 0.77$,
    $\Delta y_b =  0.45$,  and:
    $H=1.27$ (blue solid circles),
    $H=1.17$ (green open circles), and
    $H=1.07$ (red open squares).
    The first  set (solid circles) has infinite horizon while the other two  have finite horizon.
    The dashed line corresponds to $\zg(\moment) = \moment/2$
    while the dotted curves are $\zg(\moment) = \moment  - \moment_c/2$ with (from top to bottom)
    $\moment_c = 2$, $3$, and $4$, respectively.
    Inset: Schematic geometry of the polygonal channel.
    \label{fig:PBC-moments}}
\end{figure}

The scatterers of the Lorentz gas are dispersive such that its dynamics is chaotic.
Drastically different are the polygonal billiard channels (PBC)~\cite{ARV02,JR06,SL06,SS06}.
Their dynamics is that of point particles undergoing specular reflections with upper and lower straight walls,
periodically arranged in one direction.
The geometry of the elementary cell is specified by four parameters
as shown in the inset of \Fig{PBC-moments},
where we follow the notation that has been used in~\cite{JR06}.
When  $\Delta y_b + \Delta y_t  <  H$ the channel has an infinite horizon,
meaning that a trajectory can move for arbitrarily large distances without colliding with a wall.

We consider in this case an ensemble of particles that start at a random position in a random direction in a unit cell of the periodic channel.
Similarly to the action of the SM, a bundle of nearby trajectories only separates when ``sliced'' by a corner.
The resulting sub-populations of the bundle will follow qualitatively different paths
after they hit separate line segments of the wall. 
Analogously to the LG case,
the probability density of trajectory segments of length $l$ scales as  $\sim   l^{-3}$.
For all PBCs there are families of trajectories
which travel arbitrarily large distances in a given direction
without reversing their motion~\cite{JR06,SL06,SS06},
effectively acting as light fronts.

In other aspects the dynamics of the PBC differ substantially from the LG.
Firstly, in the PBC the separation of trajectories produced by the channel corners is not exponentially fast.
Secondly, the ballistic excursions appear not only as the result of an infinite horizon
but also because of the poor mixing due to non-defocusing collisions~\cite{JR06,SL06}.

In \Fig{PBC-moments} we show the spectrum $\zg(\moment)$
for three different PBC with
$\Delta x = 1$,
$\Delta y_t = 0.77$,
$\Delta y_b = 0.45$ and different widths $H$ (previously considered in~\cite{SL06}).
The spectrum $\zg(\moment)$ is described by 
\begin{equation} \label{eq:spec-PBC}
  \zg(\moment) = \left\{
    \begin{array}{l@{\quad\text{for \ }}l}
                 \frac{\moment}{2}         &   \moment < \moment_c \,,\\
       \moment - \frac{\moment_c}{2}       &   \moment > \moment_c \,.
    \end{array} \right.
\end{equation}
with $\moment_c = 2$, $3$, and $4$ for $H = 1.27$, $1.17$, and $1.07$, respectively.

For  $H = 1.27$ the PBC has an infinite horizon, and we observe the same spectrum of the moments of the displacement as for the LG.
Quite interesting is the behavior of PBCs with finite horizon, 
$H=1.17$ (green open circles) and $H=1.07$ (red open squares).
All trajectories experience collisions with the walls within a maximum finite time.
However, the bundles of trajectories
that mimic light fronts have a weaker impact on the spectrum of the moments of the displacement~\cite{JR06}.
In particular they do not dominate the mean-square displacement.
Moments dominated by the light front arise for $\moment > \moment_c > 2$. 
For $H=1.17$ the ballistic scaling sets in at  $\moment_c = 3$ and for $H=1.07$ at $\moment_c = 4$.
Therefore, matching the mean-square displacement with that of the FnD dynamics does not provide information on the
spectrum of the moments of the displacement for any other value of~$\moment$.
On the other hand, for $\xi = \moment_c/2$ the FnD dynamics will describe the spectrum for $\moment > \moment_c$.

\subsection{Summary}

The large-$\moment$ branch of the spectrum of the moments of the displacement of  strong anomalous diffusion
can be matched with the FnD dynamics by an appropriate choice of its parameter $\xi$.
When the mean-square displacement falls into this branch, \ie~when $2 \geq \moment_c$,
the matching is obtained by equating the exponent $\zMSD$ of the mean-square displacement of the considered model
with $2-\xi$ for the FnD dynamics,
and observing that the large $\moment$ branch of $\zg(\moment)$ has slope one for systems with light fronts.

\section{Displacement autocorrelations}
\label{sec:correlations}

In this section, we discuss the scaling form, \Eq{scaling-form}, of the displacement autocorrelation functions.
We verify that the scaling holds for the analytical expressions known for the SM~\cite{GRTV17} and the LW~\cite{FB13b}.
Subsequently, we analyze the autocorrelation functions for the LLg, the LG, and the PBC,
where the correlation functions are not analytically known.
A data collapse based on \Eq{scaling-form} works in all cases.
For small $h/t$ the systems even follow the functional form of the function $C(h/t)$ of the FnD,
up to system-specific corrections for the billiard systems.
For large $h/t$ the functions $C(h/t)$ for the different systems may differ. 
In all cases the comparison to the expression \Eq{FnD-correlation-scalingForm} for the FnD provides valuable insights into the scaling of the autocorrelations.

\subsection{The Slicer Map (SM)}
\label{sec:SM-correlations}

The autocorrelation function of the SM,
$\phi(m, n) =  \left\langle (x_m - x_0) \, (x_n  - x_0) \right\rangle$,
was obtained in~\cite{GRTV17}.
In the limit of large times $m$ and $n$ one has
\begin{align}
  \phi(m, n)
  &\sim  \ \frac{ 2 }{1-\alpha} \; m \; n^{1-\alpha}
    -     \frac{ 2\, \alpha }{(1-\alpha) \, (2-\alpha)} \; m^{2-\alpha}
    + \text{const} 
\label{eq:SM-v-correlation}
\end{align}
For $m=n$, the expression reduces to the mean-square displacement:
\begin{align*}
  \phi(m, m) 
  & = \left\langle ( x_m - x_0)^2 \right\rangle
  \\
  & \sim \ \frac{4}{2-\alpha} \; m^{2-\alpha} + \frac{2}{(\alpha-1) \, (\alpha-2)} \; 2^{2/\alpha} \, .
\end{align*}

The autocorrelation function of the SM,
\Eq{SM-v-correlation},
exhibits the same scaling as \Eq{FnD-correlation-scalingForm}
after identifying $\zMSD=2-\alpha$,
as established for the spectrum of moments in \Eq{SM-moments}.
The constant terms are sub-dominant for $\eta>0$ and $n > m \gg 1$. 
Hence, we find
\begin{align}
  \frac{ \phi(m, n) }{ \phi(m,m) } - 1
  &= \frac{1}{2} \: \frac{ \zMSD }{\zMSD-1} \;
    \left[ \left( 1 + \frac{n-m}{m} \right)^{\zMSD-1} - 1 \right]  \, .
\label{eq:SM-scaling}
\end{align}
Taking  $m=t_1$  and  $n=t_2$, this expression coincides with
\Eq{FnD-correlation-scaling} for the FnD dynamics.

\begin{figure}
  \includegraphics[width=0.47\textwidth]{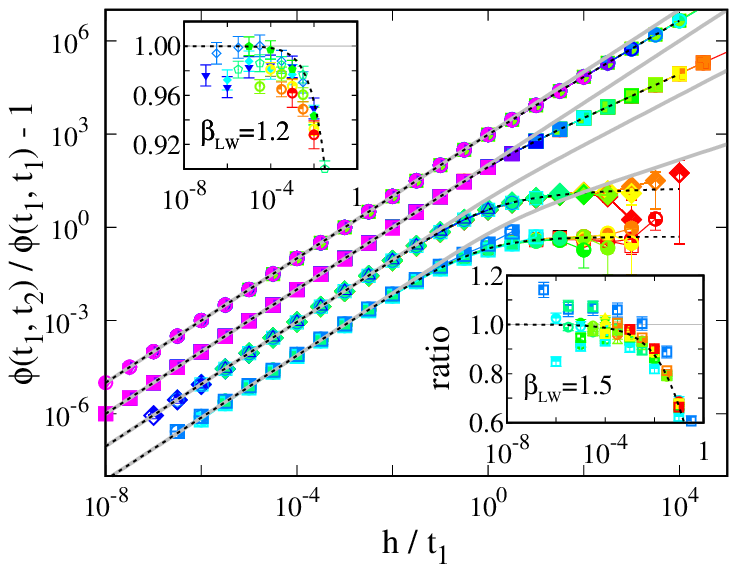}
  \caption{(Color  online).
    Scaling plot of numerical data for the autocorrelation function
    $ \phi(t_1, t_2) $
    of the LW for 
    $\zLW  = 0.1$,
    $\zLW  = 0.3$,
    $\zLW  = 1.2$,    
    and
    $\zLW =1.5$ (from top to bottom). 
    The curves are successively shifted upwards by a decade for better visibility.
    The predictions, \Eq{FnD-correlation-scaling}, are shown as gray solid lines,
    and the curves of the LW result,~\Eqs{LW-correlation-ballistic}, by black dotted lines.
    Different shades of colors indicate different choices of $t_1$,
    ranging from red for $t_1=10^3$ to violet for $t_1=10^8$.
    The symbols refer to specific choices of the time difference $h=t_2-t_1=10^{n/3}$, as follows
    $n\in\{0,12\}$~($+$),
    $\{ 1,13\}$~($\times$), 
    $\{ 2,14\}$~($\ast$),
    $\{ 3,15\}$~($\boxdot$), 
    $\{ 4,16\}$~($\blacksquare$);
    $\{ 5,17\}$~($\odot$),
    $\{ 6,18\}$~($\CIRCLE$),
    $\{ 7,19\}$~($\triangle$),
    $\{ 8,20\}$~({\large $\blacktriangle$}),
    $\{ 9,21\}$~({\large $\triangledown$}),
    $\{10,22\}$~({\large $\blacktriangledown$}),
    $\{11,23\}$~({\large $\Diamond$}).
    The insets show the ratio of the data and the prediction for
    $\zLW = 0.1$  (top left) and
    $\zLW = 1.5$  (bottom right),
    \ie~the two outermost curves.
    For better visibility of the symbols the insets only show data where $h$ is a power of ten.
    \label{fig:LW-correlations}}
\end{figure}
\subsection{L\'evy Walks (LW)}
\label{sec:LW-correlations}

In \Fig{LW-correlations} we show numerical data for the correlation functions of the LW
at times $t_1 = 10^{m/2}$ and $t_2 = t_1 + 10^{n/2}$ with $6 \leq m,n \leq N$
for (from top to bottom)
$\zLW = 0.1$ ($N=16$),
$\zLW = 0.3$ ($N=16$),
$\zLW = 1.2$ ($N=14$), and
$\zLW = 1.5$ ($N=12$).
For better visibility of the small $h/t_1$ regimes,\
the data for different $\zLW$ are shifted by successive factors of ten.
The predictions, \Eq{FnD-correlation-scaling}, are shown as gray solid lines.
The value of the time $t_1$ is indicated by rainbow colors ranging from red for $t_1=10^3$ to violet for $t_1=10^8$.
The time difference $h$ is indicated by symbols, as specified in the figure caption.
The FnD predictions, \Eq{FnD-correlation-scaling}, are shown as thick gray solid lines.
The dotted black curves show the scaling function derived from the exact solution of Froemberg and Barkai~\cite{FB13b} (details are given below).
There is a perfect data collapse for all values of~$\zLW$.
For $h/t_1 \lesssim 10^{-2}$ there even is quantitative agreement between the LW and the FnD prediction, \Eq{FnD-correlation-scaling}.

For $\zLW = 0.1$ we followed $2\times 10^6$  trajectories for a time~$10^{9}$, and 
for $\zLW = 0.3$ we followed $5 \times 10^5$ trajectories to the same time $10^{9}$.
This provides data for $N\leq 16$ that allows us to explore the correlation function
in the range of reduced times, $10^{-8} \leq h/t_1 \leq 10^{5}$,
\ie~for $13$ orders of magnitude.
For all values $\zLW < 1$ the LW shows ballistic transport with exponent $\zMSD = 2$
such that \Eq{FnD-correlation-scaling} reduces to the linear function
$C(h/t_1) = h/t_1$. 
For $h/t_1 \lesssim 1$ the data follow the predicted linear scaling. 
However, for larger $h/t_1$ it is better described by power laws with exponents $0.9$ and $0.7$, respectively (not shown).
In this range the data characterize the decay of correlations.

Froemberg and Barkai~\cite{FB13b} showed
that in the ballistic regime the correlation function involves a sum of three incomplete Bessel functions of $t_1/t_2$
with parameters that depend on $\zMSD$ and prefactors proportional to $t_1\,t_2$, $t_2^2$, and $t_1^2$, respectively.
This entails the scaling form, \Eq{scaling-form}, since $\phi(t_1,t_1) \sim t_1^2$.
Moreover, in their Eq.~(15) they provide the asymptotic scaling of $\phi(t_1, t_1+h)$ for small and large $h/t_1$.
In combination with the mean-squared displacement their result provides for $\zMSD = 2$ 
\begin{subequations}\label{eq:LW-correlation-ballistic}
\begin{align} 
    C_{LW}\!\left( \frac{h}{t_1} \right) = \left\{\!
    \begin{array}{lll}
      \frac{h}{t_1}
      &\text{ for \ }
      & \frac{h}{t_1} \ll 1 \, , 
      \\[2mm]
      \frac{ \sin( \pi \, \zLW) }{\pi\, \zLW \: (1-\zLW^2) } \; \left(\frac{h}{t_1}\right)^{1-\zLW}
      &\text{ for \ }
      &\frac{h}{t_1} \gg 1 \, . 
    \end{array} \right.
\end{align}
For $h/t_1 \gg 1$ this results predicts the observed crossover to power laws with exponent $1-\zLW=0.9$ and $0.7$, respectively.
This exponent depends on $\zLW$, and clearly FnD tells nothing about this crossover.
However, for $h/t_1 \lesssim 1$ the functional dependences agree.
This agreement is remarkable because for ballistic transport the LW shows single-scale ballistic scaling.
There is not even strong anomalous diffusion.

For $\zLW = 1.2$ we followed $1.2\times 10^5$ trajectories to a time $10^{8}$,
providing data covering $11$ orders of magnitude,  $10^{-7} \leq h/t_1 \leq  10^{4}$.
\EQ{LW-moments} provides $\zMSD(\zLW = 1.2) = 1.8$ and accordingly 
\Eq{FnD-correlation-asymptotics} predicts a crossover
from a power law with exponent $1$ for $h/t_1 \ll 1$
to an exponent $0.8$ for  $h/t_1 \gg 1$.
The upper left inset of \Fig{LW-correlations} shows the ratio of the data and the FnD prediction.
For small $h/t_1$ the data approach the FnD asymptotics with deviations of only a few percent
that lie within the scatter of our data. 
For large $h/t_1$ the data lie below the FnD dependence.
In the main panel one sees that for $h/t_1 \gtrsim 10$ the function is constant to within our numerical resolution.

For $\zLW = 1.5$ the numerical simulations are still more challenging:
we only followed $9 \times 10^4$ trajectories till time  $10^{7}$,
providing data for $10^{-6} \leq h/t_1 \leq  10^{5}$.
For $h/t_1 \lesssim 10^{-2}$ they follow the FnD prediction with a scatter of about $20$\%,
as demonstrated in the lower right inset of \Fig{LW-correlations}.
Also these data fall below the FnD dependence for larger values of $h/t_1$,
taking approximately constant values for $h/t_1 \gtrsim 1$.

According to Ref.~\cite{FB13b} the correlation function of a LW with length and velocity scales $r_0 = v =1$
takes the following form in the superdiffusive regime 
\begin{widetext}
  \begin{align*}
    \phi(t_1, t_1+h) = \frac{\zLW-1}{\zLW} \: \left\lvert \frac{\Gamma(1-\zLW)}{\Gamma(4-\zLW)} \right\rvert \; t_1^{3-\zLW} \: \left[
    \zLW - \left( \frac{h}{t_1} \right)^{3-\zLW} + \left( 1+\frac{h}{t_1} \right)^{3-\zLW}
    + (\zLW - 3) \: \left( 1+\frac{h}{t_1} \right)^{2-\zLW} \right]
  \end{align*}
  After substituting  $\zMSD = 3-\zLW$  this provides the following scaling form for the LW correlation function
  \begin{align}
      C_{LW}\!\left( \frac{h}{t_1} \right) = \frac{ \zMSD - 1 - \left( \frac{h}{t_1} \right)^{\zMSD} + \left( 1+\frac{h}{t_1} \right)^{\zMSD}
      - \zMSD \: \left( 1+\frac{h}{t_1} \right)^{\zMSD-1} }{ 2 \: (2-\zMSD) }
      = \left\{ \begin{array}{lll}
                  \frac{\zMSD}{2} \; \frac{h}{t_1}
                  & \quad\text{for \ }
                  &\frac{h}{t_1} \ll 1 \, , \; 1 < \zMSD < 2 \, ,
          \\[2mm]
                  \frac{ \zMSD -1 }{ 2 \: (2-\zMSD) }
                  & \quad\text{for \ }
                  &\frac{h}{t_1} \gg 1 \, , \; 1 < \zMSD < 2 \, .
        \end{array} \right. 
  \end{align}
\end{widetext}
For small $h/t_1$ this is identical to the FnD expression, \Eq{FnD-correlation-scaling},
and for large $h/t_1$ it accounts for the observed saturation of the scaling function. 
\end{subequations}

We conclude that an important qualitative prediction of the FnD model holds for all data sets:
the displacement autocorrelation function of the LW
admits a data collapse in the form of \Eq{scaling-form}.
Moreover, the FnD provides an accurate, parameter-free description for small values of the reduced time~$h/t_1$.
On the other hand, for $h/t_1 \gg 1$ the functional dependence of the reduced correlation functions for FnD and the LW differ.
Indeed, for a strongly ergodic system the correlation
$\phi(t_1, t_2) = \langle x(t_1)\, x(t_2)\rangle$
should factorize in the limit of $t_1, t_2, h/t_1 \to \infty$
and decay to zero for a symmetric dynamics,
$\langle x(t_1) x(t_2) \rangle \to  \langle x(t_1) \rangle\langle x(t_2) \rangle = 0$.
As a consequence, $\phi(t_1,t_2) / \phi(t_1, t_1) -1 \to -1$ for large $h/t_1$.
However, neither the FnD nor the LW enjoy such ergodic features \cite{MJCB14}.

\subsection{The L\'evy-Lorentz Gas (LLg)}
\label{sec:LLg-correlations}

\begin{figure}
  \includegraphics[width=0.47\textwidth]{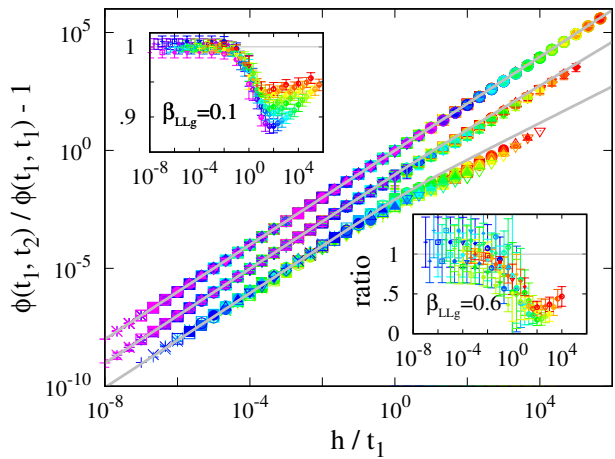}
  \caption{(Color  online).
    Scaling plot of numerical data for the autocorrelation function
    $ \phi(t_1, t_2) $
    of the LLg for 
    $\zLLg  = 0.1$ (upper curve),
    $\zLLg  = 0.3$ (middle curve),
    and
    $\zLLg =0.6$ (bottom curve), respectively. 
    Successive curves are shifted vertically by a decade for better visibility.
    The predictions, \Eq{FnD-correlation-scaling}, are shown as gray solid lines.
    The choice of colors and symbols matches with those in \Fig{LW-correlations}. 
    The insets show the ratio of the data and the prediction
    for $\zLLg=0.1$  (top left)  and
    for $\zLLg=0.6$  (bottom right), respectively.
    Also here we only show data where $h$ is a power of ten.
    \label{fig:LLg-correlations}}
\end{figure}

For the LLg comprehensive numerical simulations for the autocorrelation function, \Eq{phiDef},
have been reported in~\cite{GRTV17}.
For the present study we performed additional simulations
where we evaluated the correlation functions for the times
$t_1 = 10^{m/3}$ and $t_2 = t_1 + 10^{n/3}$ with $9 \leq m,n \leq N$.
\Fig{LLg-correlations} shows the scaling representation, \Eq{FnD-correlation-scaling},
for three different parameter values of the LLg,
$\zLLg = 0.1$ (upper  curve, $N = 24$),
$\zLLg = 0.3$ (middle curve, $N = 24$), and
$\zLLg = 0.6$ (lower  curve, $N = 21$), respectively.
The value of the time $t_1$ and $h$ are indicated by the same choice of rainbow colors
and symbols as in \Fig{LW-correlations}.
Again the data for different $\zLLg$ are shifted by successive factors of ten.

As suggested by \Eq{scaling-form}
all correlations collapse in a master curve
when $\phi(t_1,t_2)/\phi(t_1,t_1) - 1$ is plotted as function of  $h/t_1$.
The FnD expression for the asymptotic scaling, \Eq{FnD-correlation-scaling},
is provided by solid gray lines. 

For $\zLLg = 0.1$ we performed simulations where we followed $10^3$ trajectories in each of $2200$ realizations of the positions of the scatterers for times up to $10^{10}$. 
This allows us to explore the correlation function over $14$ orders of magnitude 
in the reduced times, $10^{-8} \leq h/t_1 \leq 10^{6}$. 
The mean-square displacement grows like $t^{1.991}$.
Consequently, the slopes of the small and large $h/t_1$ branches of the scaling function are almost identical,
$1$ and $\zMSD-1 = 0.991$, respectively. 
The data lie right on the prediction over the full parameter range of~$h/t_1$.
Our estimate of the statistical errors of the data are shown by error bars.
They are smaller than the data points in this figure. 
No parameter is adjusted to achieve this remarkable agreement.

For increasing values of $\zLLg$ the trajectories perform more collisions per unit time.
Hence, it becomes harder to reach large times in the simulations,
but at the same time one needs fewer realizations of disorder to achieve faithful averages.
For $\zLLg = 0.3$ we followed $10^3$ trajectories in each of $1150$ realizations of the scatterers
up to a time $10^{9}$.
This allows us to follow the correlation function
in the range of reduced times, $10^{-8} \leq h/t_1 \leq 10^{5}$. 
Estimates of the statistical errors are smaller than the points for most data.
The mean-square displacement grows with $\zMSD = 1.93$
such that the two branches have slopes $1$ and $\zMSD-1 = 0.93$, respectively. 
The change of slope is then becoming noticeable for our numerical data.
Also in this case the data lie on the prediction over the full parameter range of $h/t_1$
that is $13$ orders of magnitude wide.

For $\zLLg = 0.6$ we followed $10^3$ trajectories in each of $260$ realizations of the scatterers
up to a time of $10^{8}$. 
This allows us to explore the correlation function
in the range of reduced times, $10^{-7} \leq h/t_1 \leq 10^{4}$,
\ie~for $11$ orders of magnitude.
The mean-square displacement grows with $\zMSD = 1.775$
such that the two branches have clearly different slopes,~$1$ and~$0.775$, respectively.
For $h/t_1 \lesssim 1$ the data follow exactly the predicted linear scaling.
However, for larger $h/t_1$ they tend to lie slightly below the prediction.

In the inset to the upper left we show the ratio of the data for $\zLLg = 0.1$
and the FnD prediction.
There is a perfect match for $h/t_1 \lesssim 10^{-2}$,
and the data deviate systematically towards smaller values for larger~$h/t_1$.
However, the deviation is smaller when both $t_1$ and $h$ are large.
The same is observed for $\zLLg=0.3$ with a maximum deviation of $50$\%.

The inset to the lower right shows the corresponding plot for $\zLLg = 0.6$.
Also in this case there is a very good match for $h/t_1 \lesssim 10^{-2}$,
and the data deviate systematically towards smaller values for larger~$h/t_1$.
The maximum deviation reaches a factor of five for $h/t_1 \simeq 10^2$, 
and we see the same trend, that it becomes smaller again for still larger values of $h/t_1$.

In accordance with our expectation for a correction to scaling
the deviations differ for different times $t_1$ at a fixed ratio $h/t_1$,
as indicated by the different colors of the symbols. 
In contrast, according to \Eq{FnD-correlation-ratio} 
the corrections to scaling in the FnD dynamics approach one from above when $t_1$ is increased.

For all data sets an important qualitative prediction of the FnD model holds:
the displacement autocorrelation function of the LLg 
obeys the scaling form \Eq{scaling-form}.
Moreover, the FnD expression, \Eq{FnD-correlation-scalingForm},
provides a rather accurate, parameter-free description of the two-variable correlation function $\phi(t_1, t_2)$
in terms of the time ratio~$h/t_1$.
There are deviations in particular at intermediate times.
They constitute a non-universal correction to \Eq{FnD-correlation-scaling},
and they behave qualitatively differently in the LLg and the FnD dynamics.

\subsection{The Lorentz gas (LG)}
\label{sec:LG-correlations}

\begin{figure}[t]
  \centering
  \includegraphics[width=0.47\textwidth]{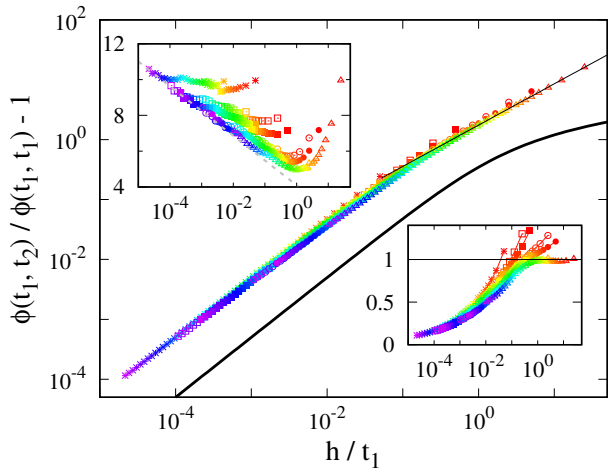}
  \caption{(Color online).
    Scaling plot for the autocorrelation function
    $\phi(t_1, t_2)$ 
    of the LG with infinite horizon $R=0.1$.
    The time $t_1$ takes values $200 \leq t_1 < 5\times 10^5$
    that are marked by rainbow colors ranging from red for $t_1=200$ till violet for $t_1 = 5\times 10^5$
    on a logarithmic scale. 
    The symbols refer to specific choices of the time difference $h=t_2-t_1$ 
    that takes the values
    $10$~($\ast$),
    $50$~($\boxdot$),
    $100$~($\blacksquare$),
    $500$~($\odot$),
    $1000$~($\CIRCLE$), or
    $5000$~($\triangle$), respectively.
    The prediction, \Eq{FnD-correlation-scaling}, with  $\zMSD = 2$ 
    is shown by a thick solid black line.
    The inset to the upper left shows the ratio of the data point and the prediction.
    The dotted gray line indicates the logarithmic law $4.1-0.6 \, \log(h/t_1)$.
    The inset to the lower right shows the ratio of the data and the power law
    $1.75 \: (h/t_1)^{0.69}$, that is shown by a thin black solid line in the main panel.
    \label{fig:LG-correlations}}
\end{figure}

We have numerically computed the autocorrelation function of the Lorentz gas
$\phi(t_1,t_2) = \langle  \Delta x(t_1)  \Delta  x(t_2) \rangle$,
where  $\langle  \cdot  \rangle$ refers to an average over an ensemble of
$1.8\times 10^8$ trajectories
that is evaluated for $200 \leq t_1 \leq 5\times 10^5$  
and different time increments $h=t_2-t_1 \in \{10, 50, 100, 500, 1000, 5000\}$.
The results for the infinite-horizon case, $R=0.1$, are shown in \Fig{LG-correlations},
where time~$t_1$ is marked by color and~$h$ by symbols, as indicated in the figure caption.
As for the LLg, the data accurately collapse to a line
when the reduced correlation function
$\phi(t_1,t_2) / \phi(t_1,t_1) - 1$ 
is plotted as function of $h/t_1$.
The mean-square displacement grows with exponent $\zMSD = 1$,
and the prediction \Eq{FnD-correlation-scaling} is provided by the thick solid black line.
The parameter-free prediction provides values that are smaller by about one order of magnitude than the observed data.

The ratio of the data and the prediction is provided in the upper left inset:
the deviation for $h/t_1 \lesssim 1$ amounts to leading order to a logarithmic law.
Encountering logarithmic corrections to scaling is not unexpected for the LG,
where they also arise in the mean-square displacement \cite{B92,ZPFDB18,FouxonDitlevsen2019}.
Likewise, we see here numerical evidence for a logarithmic contribution, $\ln(h/t_1)$, to the asymptotic scaling function.
On top of that there are noticeable finite-time corrections. 
The data for different time increments $h$ bend away from the logarithmic line,
taking a minimum towards right for $t_1 \simeq 5000$,
and they increase for larger values of $h/t_1$.
Similarly to the finite-time corrections of the FnD,
the asymptotic scaling is approached from above.

For $h/t_1 \gtrsim 1$ the FnD prediction,
that is shown by the thick solid line in the main panel,
looks qualitatively different than the numerical data.
Rather than following the $\zMSD=1$ case of \Eq{FnD-correlation-scaling}
the numerical data are better described by a power law. 
A guide to the eye is provided by the thin black line that shows the dependence $1.75 \: (h/t_1)^{0.69}$.
The ratio of the data and this power law is shown in the lower right inset.
It reveals that the fit refers only to the final decade of the $h=5000$ dependence.
Much longer simulations are required, therefore, to make qualified statements about the large $h/t_1$ asymptotics of the correlation function. 
In particular, when the correlation decays for vastly different times $t_1$ and $t_2$
we would again expect that the scaling function must eventually approach $-1$ in the large-time limit,
$1 \ll t_1 \ll t_2$.

We conclude that the important qualitative prediction, \Eq{scaling-form}, of the FnD model holds also for the LG:
in the large-time limit the ratio of the displacement autocorrelation function and the mean square displacement is a function of only $h/t_1$.
However, in this case the FnD prediction of the scaling function, \Eq{FnD-correlation-scalingForm},
provides only a rough idea of the form of the scaling function.
It misses logarithmic terms in the small $h/t_1$ limit, and for large  $h/t_1$ it is off even to leading order.

\begin{figure*}[t]
  \centering
  \[
    \includegraphics[width=0.47\textwidth]{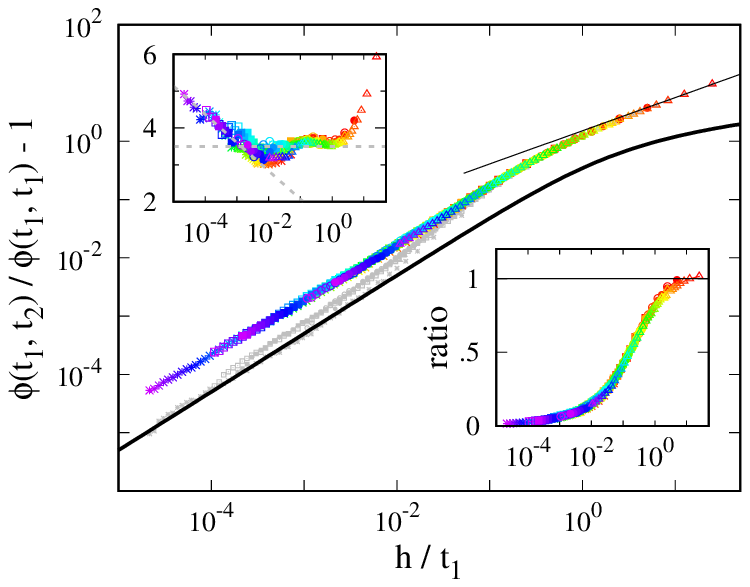}
    \quad
    \includegraphics[width=0.47\textwidth]{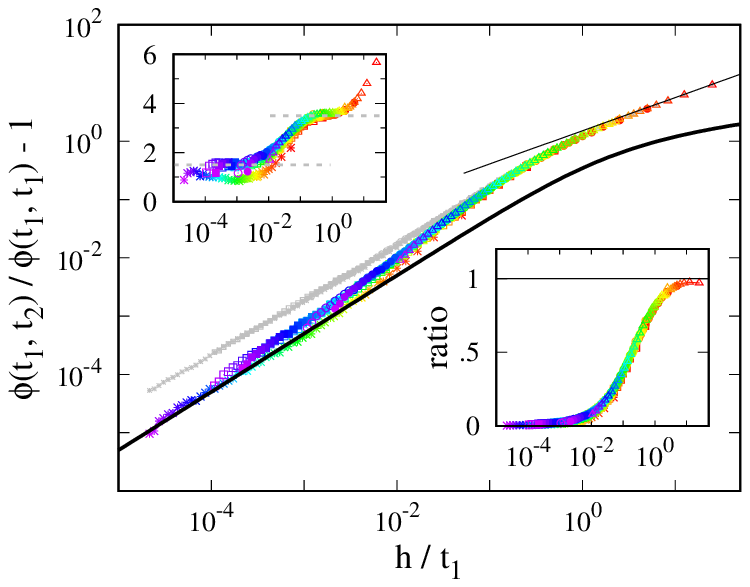}
  \]
  \caption{(Color online).
    Scaling plot for the autocorrelation function
    $\phi(t_1, t_2)$ 
    of a PBC with infinite horizon ($H=1.27$, left), and
    the with finite horizon ($H=1.07$, right).
    Data are evaluated for the same values of  $t_1$ and $h$ as for the LR,
    see \Fig{LG-correlations}.
    The prediction, \Eq{FnD-correlation-scaling}, with  $\zMSD = 2$ 
    is shown as thick solid black curve.
    The theory-curve is the same as for the LG and we also adopt the same symbols and colors 
    to indicate $h$ and $t_1$, respectively.
    The upper left insets shows the ratio of the data and the FnD prediction, \Eq{FnD-correlation-scaling}.
    The dashed gray lines show offset values and a logarithmic law, $1.3 - 0.33 \, \log(h/t_1)$,
    that are discussed in the main text.
    The lower right insets show the ratio of the data and power law
    $1.52 \: (h/t_1)^{0.57}$ 
    that provides a good fit to the large $h/t_1$ asymptotics of the data for both channels.
    The fit is shown by thin solid black lines in the main panels.
    For visual inspection the data for the other channel are indicated by smaller gray symbols in the respective plots.
    \label{fig:PBC-correlations}}
\end{figure*}
\subsection{Polygonal billiard channels (PBC)} 
\label{sec:PBC-correlations}

For PBCs we have numerically computed the autocorrelation function $\phi(t_1, t_2)$
for the polygonal channel
with infinite horizon~$H=1.27$ and
with finite horizon $H=1.07$.
The results are shown in the left ($H=1.27$) and right ($H=1.07$) panel of \Fig{PBC-correlations}.
We show data for the same combinations of $h$ and $t_1$ as adopted for the LG,
and also use the same color coding and symbols. 
The scaling form provides a data collapse also for these billiards. 

For both channels the mean-square displacement scales with $\zMSD=1$, \ie~with the same exponent as the LG.
Hence, the FnD model suggests the same scaling function for these three types of vastly different billiards.
However, the data differ for the different billiards.

The data for the infinite-horizon case, $H=1.27$,
deviate by a factor of $3.5\pm 1$ from the FnD prediction,
as indicated by the horizontal dashed gray line in the upper left inset in the left panel.
Moreover, for small $h/t_1$ the infinite horizon PBC also shows logarithmic corrections to the FnD prediction,
analogous to the correction observed for the LG with infinite horizon.
Also for large $h/t_1$ the data do not follow the logarithmic law predicted by \Eq{FnD-correlation-scalingForm},
but they are better described by a power law with an exponent $0.57$
that is shown by a thin solid black line in the main panel.
The ratio of the data and this power law is shown in the lower right inset.
The data for different times $t_1$ collapse considerably better in this case than in the LG,
and they all nicely approach the asymptotic power law.
However, also in this case much longer simulations are required to make qualified statements about the large $h/t_1$ asymptotics of the correlation function. 

For $H=1.07$ the data lie closer to the FnD prediction.
For  $h/t_1 \lesssim 10^{-2}$ there is a fixed factor of $1.5$ between the data and the FnD prediction.
This is indicated by a horizontal dashed gray line in the upper left inset,
which shows the ratio of the data and the FnD prediction.
For the PBC with a finite horizon the FnD prediction therefore provides the right scaling of the data,
and it is off in the prefactor of the scaling by $50$\%.
This indicates that the sub-leading logarithmic contribution of the scaling function for $h/t_1 \ll 1$ is indeed a property of the infinite horizon. 

In the range $0.002 < t_1 < 0.2$ the offset of the finite-horizon PBC increases to a factor of $3.5$,
that is indicated by another dashed line in the inset.
For still larger $h/t_1$ the scaling functions of both PBCs coincide.
This suggests that the decay of correlations for $h/t_1 \gg 1$ follows the same dependence for PBC with finite and infinite horizon.

We conclude that the important qualitative prediction, \Eq{scaling-form}, of the FnD model holds also for the PBC:
in the large-time limit the ratio of the displacement autocorrelation function
and the mean square displacement is a function of only $h/t_1$. 
In this case the FnD prediction of the scaling function, \Eq{FnD-correlation-scalingForm},
provides an accurate description of the small $h/t_1$ dependence of the billiards with finite horizon,
up to constant factor of $1.5$. 
This agreement is remarkable
because the FnD dynamics does not even describe the mean-square displacement of the PBC with finite horizon.
This suggests that for $h/t_1 \ll 1$ the scaling form, \Eq{FnD-correlation-scalingForm},
is a generic feature of strong anomalous diffusion.

\section{Deviations from scaling}
\label{sec:discussion}

The scaling form, \Eq{scaling-form},
of the displacement correlation function, \Eq{phiDef},
provides a data collapse for all dynamics.
For the LLg with small parameter values~$\zLLg$ we find quantitative agreement between our numerical data
and the parameter-free prediction of the FnD model over the full range of the dimensionless time difference $h/t_1$,
that is varied over
    $14$ orders of magnitude for $\zLLg=0.1$ (uppermost curve in \Fig{LLg-correlations}),
for $13$ orders of magnitude for $\zLLg=0.3$ (middle curve), and
    $11$ orders for $\zLLg=0.6$ (bottom curve).
This data closely follows the FnD prediction, 
showing systematic deviations from scaling only at intermediate values of $h/t_1$.
We attributed these deviations to corrections to the scaling prediction for finite values of $t_1$.
For the FnD model such deviations have been established in \Eq{FnD-correlation-ratio}.
These corrections for non-asymptotic regimes are non-universal,
as it may be expected.
For the FnD model finite-time data are larger than the asymptotic values,
for the LLg we find smaller values. 
For small values of $h/t_1$ the numerical data agree quantitatively with the FnD prediction,
and the quantitative difference are small over the full range of $h/t_1$:
The differences increase from $10$\% for $\zLLg=0.1$ to $60$\% for $\zLLg=0.6$.

For the LW the scaling prediction, \Eq{scaling-form}, provides a data collapse without appreciable finite-time corrections.
For  $h/t_1 \lesssim 0.01$ the FnD model even provides a quantitative description without adjustable parameters.
The upper two curves in \Fig{LW-correlations} demonstrate that this even applies
in the ballistic regime.
For large $h/t_1$ the LW data lie below the FnD prediction,
with increasing deviations for larger~$\zLW$.
Comparison with the scaling function for the LW, \Eq{LW-correlation-ballistic}, reveals
that the scaling form, \Eq{scaling-form}, holds over the full range of $h/t_1$,
and that the scaling functions of the LW and of FnD match for small $h/t_1$.

For the LG with infinite horizon the data show a collapse with a scaling function
that follows the trend predicted by FnD (upper curve in \Fig{LG-correlations}).
However, there is no quantitative agreement of the scaling functions in this case. 
For small $h/t_1$ we identified logarithmic corrections
that we expect to persist in this system (dashed line in the upper left inset).
Such corrections are in accordance with the logarithmic contribution to the scaling of the statistics of displacements \cite{B92}.
However, they take the qualitatively new form of a $\log(h/t_1)$ dependence.
For large $h/t_1$ the FnD model predicts a logarithmic dependence,
while the data are better described by a power law.
The correlation $\phi(t_1, t_2)$ does not decay to the ensemble average 
if such a power law describes the $h/t_1$ asymptotics. 
After all, due to the left-right symmetry of the LG the steady state equilibrium value of $\langle x(t_1) x(t_2)\rangle$ 
must vanish in the limit $1 \ll t_1 \ll t_2$.
The very long transients of the LG~\cite{ZPFDB18,FouxonDitlevsen2019} make it impossible for us to explore this crossover.

The PBC with infinite horizon also enjoys a faithful data collapse,
with very small scatter for data evaluated at finite times. 
All data lie in a band that differs only by a factor of $3.5\pm 1$ from the FnD prediction
(left panel in \Fig{PBC-correlations}).
Still, the deviations are systematic, of exactly the same type as for the LG.
For small $h/t_1$ the PBC with infinite horizon shows logarithmic corrections.
For large $h/t_1$ the data are better described by a power law than by the logarithmic dependence predicted by the FnD model. 
The data for the PBC with finite horizon also show a faithful data collapse,
with noticeable deviations from the master curve only for the smaller time increment $h=10$.
For $h/t_1 \lesssim 10^{-2}$ the data follow the FnD prediction up to a factor~$1.5$.
Subsequently, they cross over to the dependence also observed for the PBC with infinite horizon. 
This agreement for large values of $h/t_1$ indicates a common mechanism for the decay of correlations in the PBC,
this is characterized by the autocorrelation function for $t_2 \gg t_1 \gg 1$.

It is remarkable that the FnD model does not only suggest a scaling form, \Eq{scaling-form}, for the displacement correlation function
that is followed by all models investigated here,
but that there even is quantitative agreement between the predicted scaling function
and the numerical data for the LW, LLg and the PBC with finite horizon. 
We suggest that this can be explained based on the Buckingham-Pi theorem~\cite{Buck14,Barenblatt}.
To this end we consider the following choices of scales and dimensionless parameters:
\begin{description}
\item[FnD] 
  The time scale is set by the minimum flight time~$t_M=b^{1/\xi}$, \Eq{tcx0} and below,
  while space and time scales are related by the unit velocity of the particles.
  There are no dimensionless groups in this model. 

\item[SM]
  The length scale is set by the box size,
  while space and time scales are related by the unit velocity of the particles.
  There are no dimensionless groups in this model.

\item[LW]
  The length scale is set by the minimum length of trajectory segments,
  while space and time scales are related by the unit velocity of the particles.
  There are no dimensionless groups in this model.

\item[LLg]
  The length scale is set by the minimum distance of scatterers,
  while space and time scales are related by the unit velocity of the particles.
  There are no dimensionless groups in this model.

\item[LG]
  The length scale is set by the scatterer separation $\Delta$
  while space and time scales are related by the unit velocity of the particles.
  The ratio of scatterer distance and scatterer radius provides a dimensionless parameter $\Delta / R$
  (\cf~the inset of \Fig{LG-moments}).

\item[PBC]
  The length scale is set by the periodicity $2\,\Delta x$ of the channel,
  while space and time scales are related by the unit velocity of the particles.
  Three dimensionless groups are required to fully characterize the channel,
  $\Delta y_t / \Delta x$, $\Delta y_b / \Delta x$, and $H / \Delta x$
  (\cf~the inset of \Fig{PBC-moments}).
\end{description}
For small $h/t_1$ the FnD model provides a prediction, \Eq{FnD-correlation-scalingForm}, on the exponents,
while the prefactors of the power laws will in general depend
on the dimensionless groups of the models~\cite{Buck14,Barenblatt}. 
From this perspective, we expect full agreement of the FnD, SM, LW, and the LLg,
while the prefactors of the power laws are expected to depend on the system geometry for the LG and PBCs.
The scaling behavior for large $h/t_1$ is not expected to be universal. 
It will depend on the relaxation to the non-equilibrium distribution,
and may suffer from breaking of ergodicity and persistent transients.

\section{Conclusions}
\label{sec:conclusion}

The moments of the displacement of many systems
that show strongly anomalous superdiffusive transport
are dominated by ballistic trajectories or light fronts~\cite{CMMGA99,VBB18,LastBurioniArXiv}.
Often they lead to a two-piece linear scaling of the exponent $\zg(\moment)$
of the temporal growth of the~$p^{\text{th}}$ moment of the probability distribution of the displacement, \Eq{SAD}.
Here, we presented the very simple Fly-and-Die (FnD)  model 
that arguably is the simplest dynamics in this class of systems.
Due to the simplicity of its dynamics one can derive 
analytical expressions for the exponents $\zg(\moment)$
and the displacement autocorrelation function $\phi(t_1, t_2)$.
Based on the FnD model we have proven
that the correlation function follows a universal scaling, \Eq{scaling-form}, for a range of systems showing strong anomalous diffusion.
For {\emph{all} system
where the second moment is governed by a light front
the ratio $\phi(t_1, t_2)/\phi(t_1,t_1)$ is a function $1+C(h/t_1)$ of a single argument,
$h/t_1 = (t_2-t_1)/t_1$ with $t_2>t_1$.
We thus relate the scalings of the position-position displacement correlation function
and of the mean-square displacement,
connecting the most widely studied and best characterized property of these systems
to a feature that is not known for most strongly anomalous systems.
The prediction applies for all systems mentioned in the Introduction.

In the second part of the manuscript
we underpinned our claim by discussing
representative models from five classes of widely studied systems:
i.~the Slicer Map, a deterministic, time-discrete dynamical system,
ii.~L\'evy walks, where strong anomalous diffusion arises from following a route of straight line segments whose length are sampled from a distribution with a power-law tail~\cite{MJCB14,ZDK15}, 
iii.~the L\'evy-Lorentz gas, a random-walk model featuring quenched disorder~\cite{BCV10}, 
iv.~the Lorentz gas, a chaotic billiard with infinite horizon~\cite{AHO03}, and
v.~polygonal billiard channels, where trajectories only separate when they hit different walls of the polygon~\cite{JR06}.
For the billiards we considered systems with finite and infinite horizon.

The FnD model makes two predictions that are robust
in the sense that they apply universally to all systems:

1. When the offset value $(1-\nu)\,\moment_c$ in \Eq{SAD} is fixed for one of the moments in the large $\moment$ regime,
then the scaling for all higher moments follows without further adjustable parameters.
Microscopic details of the dynamics are reflected solely by different values of $\nu$.
Specifically,
$\nu=0$ for the FnD dynamics and the SM,
$\nu=1/2$ for the billiard systems,
and $\nu$ depends on the parameter $\zLW$ for the LW
and $\zLLg$ for the LLg (\cf~\Eqs{LW-moments} and~\eqref{eq:LLg-moments}). 

2. For the FnD dynamics we derived a scaling form, \Eq{FnD-correlation-scalingForm},
of the displacement correlation function, \Eq{phiDef}.
It applies also to the SM, whose correlation function can also be calculated analytically.
Due to the peculiarities of the FnD and the SM one can not expect
that the functional form of \Eq{FnD-correlation-scalingForm} is universal.
However, it suggests that in general the autocorrelation functions admits a data collapse of the form \Eq{scaling-form}.
This is indeed true for the LW,
that features the predicted scaling (\cf~\Fig{LW-correlations}),
but with a different functional form, \Eqs{LW-correlation-ballistic}.
In addition to the LW 
we tested the data collapse with numerical data for
the LLg model with data spanning to the least eleven orders of magnitude in the reduced dimensionless time (\Fig{LLg-correlations}),
and more than six orders of magnitude for the LG with open horizon and PBCs with open and closed horizon
(\Fig{PBC-correlations}). 
For all investigated models the suggested representation of the data provides a data collapse.
For the LW and LLg the FnD prediction
even provides a parameter-free prediction of the prefactors of the asymptotic scaling laws
for small $h/t_1$. 
For the billiards the FnD prediction faithfully describes the trends and the order of magnitude of the correlation function.
In Sec.~\ref{sec:discussion} we related the offset to the presence of dimensionless parameters in the billiards
that affect the prefactors of the predicted power-law dependences.
Special care is needed in situations where the cross over, $\moment_c$, of the branches of the spectrum $\zg(\moment)$
arises at the exponent $\moment = \moment_c = 2$ of the mean-square displacement.
This is the case for billiards with infinite horizon.
For the LG and PBC with infinite horizon we observed logarithmic corrections to the scaling prediction of the FnD dynamics
that emerge for small $h/t_1$.
Moreover, the case $\zMSD = \eta(2) = 1$ is a critical dimension
where the FnD takes a logarithmic dependence for large $h/t_1$.
All considered billiard systems fall into this class,
  but their scaling in this range can better be fitted by a power law than by a logarithmic dependence.
More analytical and numerical work will be needed to fully understand these dependencies.
They will be addressed in forthcoming work that focuses on the intriguing features of billiard systems,
rather than addressing universal features of correlation functions.

3. In \Eq{FnD-correlation-ratio} we also provided the small $t_1$ corrections to scaling for the FnD model.
These corrections take a non-universal, system-specific form.
For the FnD model the asymptotic theory is approached from above.
The LW data show no noticeable finite time corrections. 
The LLg data show deviations towards smaller values.
For the LG there are noticeable deviations towards larger values.
The PBC show deviations towards smaller values only for the smallest considered time difference, $h=10$.

We conclude that FnD captures the essential features of strong anomalous transport with two underlying scales.
Only the presence of a light front and a much slower dynamics are kept, and all surplus features are removed.
In the spirit of normal forms in mathematical models our model only accounts for a single long jump. 
Its simplicity entails a straightforward analytical solution.
The present paper scrutinized the resulting predictions for the moments of the displacement and the displacement autocorrelation function.
Vastly different dynamics follow the FnD predictions for the high moments of the displacement,
\Eq{spectrum} for large $\moment$ where $\zg(\moment)$ has slope one,
and the long-time asymptotics of the displacement autocorrelation function,
\Eq{phiDef} for times $1 \ll t_2 \lesssim 2\,t_1 \leq 2 t_2$.
Moreover, the FnD establishes a scaling form for the correlations, \Eq{scaling-form}.
In the long-time regime, $1 \ll t_1 \ll t_2$,
it provides a data collapse for all investigated models,
with model-specific scaling functions for $t_2 \gtrsim t_1$. 
In the opposite limit $t_2 -t_1 \lesssim t_1 \leq t_2$ the moments and the correlations are governed by the light front.
As a consequence, the parameter-free FnD prediction is followed quantitatively, except that prefactors of scaling laws may differ:
length and time scales shift if and only if a model has dimensionless groups not addressed by the normal form.
This finding applies to all system with strong anomalous diffusion
where the second moment is governed by a light front,
and it thus calls for experimental verification in a vast range of different physical settings.

\begin{acknowledgments}
 
  JV~is grateful for the appointment as a Distinguished Visiting Professor
  at the Department of Mathematical Sciences of the Politecnico di Torino
  during the academic year 2016/17.
  LR~and MT~gratefully acknowledge computational resources provided by HPC@POLITO,
  the project for Academic Computing of the Department of Control and Computer Engineering
  at the Politecnico di Torino \url{(http://hpc.polito.it)}.
  LR acknowledges that the present research has been partially supported by MIUR grant Dipartimenti di Eccellenza 2018-2022 (E11G18000350001).
  MT~is also thankful to JV for his kind hospitality
  at the Institut f\"ur Theoretische Physik of the Universit\"at Leipzig, Germany.
  CG~acknowledges financial support form ``Fondo di Ateneo per la ricerca 2016''
  under the project
  ``Sistemi stocastici e deterministici su strutture spaziali discrete, grafi e loro propriet\'a strutturali''.
  CMM~thanks the
  Department of Mathematical Sciences of Politecnico di Torino and
  Bar-Ilan University for their hospitality,
  and to Eli Barkai for enlightening discussions.
  CMM does also acknowledges financial support from the
  Spanish Government grant PGC2018-099944-B-I00 (MCIU/AEI/FEDER, UE).

\end{acknowledgments}

\end{document}